\documentclass[preprint,11pt]{elsarticle}

\usepackage{graphicx}

\usepackage[amssymb]{SIunits}
\usepackage{latexsym}
\usepackage{marvosym}
\usepackage[usenames,dvipsnames]{color}
\usepackage{layouts}

\biboptions{sort&compress}

\newcommand{\onlyunit}[1]{\unit{}{#1}}

\newcommand{\txtsuperscript}[1]{\ensuremath{^{\textnormal{#1}}}}

\newcommand{\tttc}{T\txtsuperscript{3}C}

\journal{European Journal of Mechanics - B/Fluids}

\begin{document}

\begin{frontmatter}
\title{\normalsize{Applying Laser Doppler Anemometry inside a Taylor-Couette geometry} \\ \small{Using a ray-tracer to correct for curvature effects}}

\author[ut]{Sander G. Huisman}
\ead{s.g.huisman@utwente.nl}
\author[ut]{Dennis P.M. van Gils}
\author[ut]{Chao Sun}
\ead{c.sun@utwente.nl}
\address[ut]{Physics of Fluids, Faculty of Science and Technology\\ Burgers Center for fluid dynamics\\ University of Twente, The Netherlands \\}

\begin{abstract}
In the present work it will be shown how the curvature of the outer cylinder affects Laser Doppler anemometry measurements inside a Taylor-Couette apparatus. The measurement position and the measured velocity are altered by curved surfaces. Conventional methods for curvature correction are not applicable to our setup, and it will be shown how a ray-tracer can be used to solve this complication.

By using a ray-tracer the focal position can be calculated, and the velocity can be corrected. The results of the ray-tracer are verified by measuring an a priori known velocity field, and after applying refractive corrections good agreement with theoretical predictions are found. The methods described in this paper are applied to measure the azimuthal velocity profiles in high Reynolds number Taylor-Couette flow for the case of outer cylinder rotation.
\end{abstract}

\end{frontmatter}

\section{Introduction}
\begin{figure}[ht!]
 \begin{center}
  \includegraphics{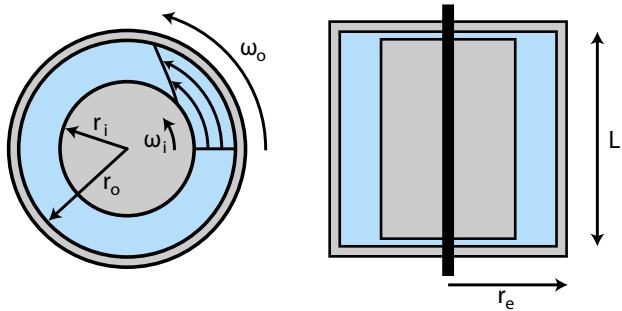}
  \caption{\emph{Left:} Top view of TC apparatus, two concentric cylinders are rotating. Control parameters are the rotation rates $\omega_i$ and $\omega_o$, where the subscripts denote inner cylinder and outer cylinder, respectively. The inner cylinder has a radius of $r_i = \unit{20}{\centi \meter}$, and the outer cylinder has an inner-radius of $r_o = \unit{28}{\centi \meter}$. \emph{Right:} Vertical cross section of a TC apparatus. The outer cylinder has an outer-radius of $r_e = \unit{30.5}{\centi \meter}$ and a height $L = \unit{92.7}{\centi \meter}$. The outer cylinder is made from optically transparent PMMA (\emph{Poly-(methyl methacrylate)}), and is attached to the top and bottom end plates.}
  \label{fig:TC_Geometry}
 \end{center} 
\end{figure}

A Taylor-Couette (TC) apparatus consists of two coaxial, differentially rotating, cylinders, see figure \ref{fig:TC_Geometry}. The annulus between the cylinders is filled with a working fluid; most commonly, as in our case, water is chosen. The apparatus has been used to study hydrodynamic instabilities, pattern formation, turbulence, and was found to have a rich phase diagram with different types of flow structures \cite{ray17,tay23,pfi81,pri81,and86,dom86,mul87,pfi88,lat92a,buc96,ess96}. To get a deep understanding of these phenomena it is crucial to measure the local flow velocity.

Measuring the velocity field inside a TC apparatus was done for a long time using intrusive measurements techniques, \emph{e.g.} constant temperature anemometry \cite{col65a,smi82,lew96phd,lew99} and Pitot tubes \cite{wen33}. Though these techniques are robust and proven to work, they are not ideal for measuring the velocity in TC flow. The aforementioned methods measure the magnitude of the velocity, not the individual components, and are directionally ambiguous using a single probe. Of course, one could use multiple probes \cite{Wallace2010} to obtain the flow direction. Another problem is that they alter the flow under consideration. Though this is not an issue for non-recirculating setups, like an open-ended wind tunnel, it can be a severe issue in recirculating (closed) setups, \emph{e.g.} a TC apparatus, a rotating drum, or a Rayleigh-B\'enard cell \cite{ahl09}. For a large range of Reynolds numbers it is known that vortices will be shedded \cite{King1977141} from these probes, either in form of a K\'arm\'an vortex street or as a turbulent wake, depending on the geometry and Reynolds number. These vortices can survive a full revolution, which has been observed in rotating drum experiments \cite{sun10}.

The TC setup used in the present work, the Twente Turbulent Taylor-Couette (\tttc)  \cite{gil10b,gil11,hui12}, distincts itself from other setups by many features: variable gap and radius ratio, precise temperature control, independently rotatable cylinders, and a fully optically accessible gap. The outer cylinder is constructed from \unit{2.5}{\centi \meter} thick PMMA (\emph{Poly-(methyl methacrylate)}), which enables optical measurement techniques, \emph{e.g.} Particle Tracking Velocimetry (PTV) \cite{Maas1993,oue06}, Particle Imaging Velocimetry (PIV) \cite{raffel2007,rav10}, and Laser Doppler Anemometry (LDA) \cite{yeh1964,albrecht2003}. These methods, by their very nature, will not disturb the flow under consideration. In addition these techniques are able to measure the velocity components and are directionally sensitive, such that they are capable of detecting flow reversals. The addition of seed particles is imperative for these techniques, and one should check if these particles accurately reflect the velocity of the flow, as discussed below. Additionally, particles should not change the dynamics of the flow, in particular, some particles act as a surfactant in two-phase flows \cite{Davis1966,Griffith1962,fdhila1996}.

\section{Laser Doppler anemometry}
\begin{figure}[ht!]
 \begin{center}
  \includegraphics{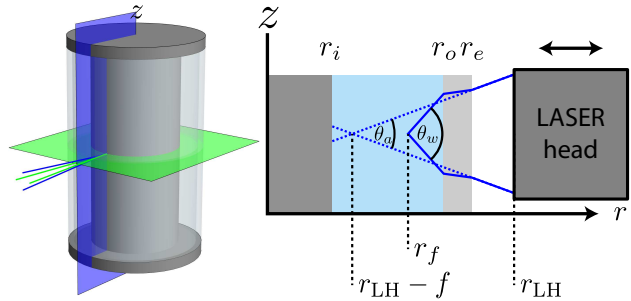}
  \caption{\emph{Left:} The azimuthal and axial components of the velocity are measured, laser beams are in the green and blue planes respectively. \emph{Right:} Vertical cross section showing two laser beams. The dashed lines are beams \emph{without} refraction, the angle between the beams is denoted $\theta_a$, where $a$ stands for air. The solid lines are beams \emph{with} refraction, $\theta_w$ is the angle between the beams in water. $r_\text{LH}$ is the position of the laser head and $r_\text{LH} - f$ is position of the focus without refraction, while $r_f$ is the real position of the focus.}
  \label{fig:PlanesAndAngle}
 \end{center} 
\end{figure}

LDA is based on the Doppler effect. The most common version of LDA, is a so-called dual beam heterodyne configuration \cite{albrecht2003}, see figures \ref{fig:PlanesAndAngle} and \ref{fig:FlatVersusCurved}. In this configuration two beams are crossed and focused in the flow, creating an interference pattern. Seed particles, added to the flow, passing through the interference pattern will scatter light with a specific frequency. This light is then captured by a photo detector and converted to a current from which the Doppler shift can be calculated. Knowing the optical geometry of the setup one can directly calculate the velocity from the Doppler shift \cite{albrecht2003}:
\begin{equation}
 f_d = \frac{2 \sin(\theta /2)}{\lambda} | v_k | \label{eq:LDAtheta}
\end{equation}
where $f_d$ is the Doppler shift, $\lambda$ the wavelength of the laser, $\theta$ the angle between the beams, and $v_k$ the component of the velocity along $\mathbf k_1 - \mathbf k_2$, where $\mathbf k_i$ are the propagation vectors of the laser beams. To add directional sensitivity one has to frequency shift one of the beams, accomplished by a Bragg cell. More details about the Bragg cell, the fringe-model, and LDA in general, can be found in \emph{e.g.} refs. \ \cite{albrecht2003,drain1980laser}.

Laser Doppler anemometry is a so-called absolute measurement method and therefore does not require calibration against a known flow. This, however, does not mean that a measurement of velocity is error free. Any misalignment in the optical arrangement, and any imperfection in the lenses (\emph{e.g.} astigmatism \cite{zhangz96}) will cause errors. In addition, any particle travelling through the beams prior to focussing can have adverse effects on the formation of a well-defined measurement volume. Similarly, any spatial inhomogeneity of the refractive index causes the focal point to shift, and the waists to mismatch in the measurement volume \cite{resagkc03}. Furthermore, the number of particles in the interference zone fluctuates; particles move in and out the measurement volume and induce noise in the collected signal. 

\subsection{Curvature effects}
\begin{figure}[ht!]
 \begin{center}
  \includegraphics{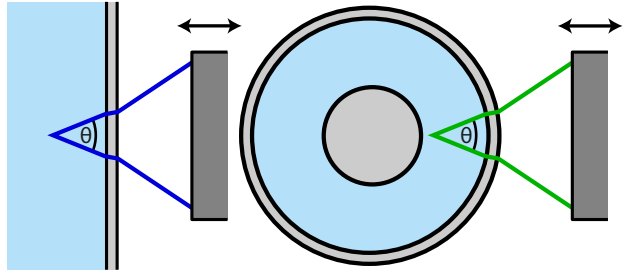}
  \caption{\emph{Left:} Typical geometry of LDA, equivalent to the vertical plane in the current application. The beams are passing through flat interfaces, and $\theta$ does not vary with laser-head position. \emph{Right:} Horizontal plane: laser beams are affected by the curved interfaces, and therefore $\theta$ is a function of radial position.}
  \label{fig:FlatVersusCurved}
 \end{center}
\end{figure}

In most LDA applications the laser beams travel through flat surfaces, see figure \ref{fig:FlatVersusCurved}. In this case, Eq. \ref{eq:LDAtheta} can be simplified by invoking Snell's law:
\begin{equation}
 \frac{f_d}{2 |v_k|} = \frac{sin(\theta_{\text{w}}/2)}{\lambda_{\text{w}}} = \frac{sin(\theta_{\text{a}}/2)}{\lambda_{\text{a}}} \label{eq:LDAsimplify} 
\end{equation}
where quantities with $a$ subscripts denote quantities in air, and $w$ in water. Equation \ref{eq:LDAsimplify} is only applicable if the interfaces are flat and the optical axis is perpendicular to those interfaces; it is only then that $\theta_{\text{a}}/2$ is the angle of incidence and $\theta_{\text{w}}/2$ the angle of refraction. The difference in refractive index is absorbed by the changing wavelength. So for the case of flat interfaces, $\theta_a$ can be obtained from the focal length and the beam separation, and together with $\lambda_a$, given by the laser, the velocity can be calculated from the Doppler shift (Eq. \ref{eq:LDAsimplify}). Note that the refractive indices of the container and water are irrelevant; they are not used in the calculation of the velocity. 

For the case of a curved surface, see Fig. \ref{fig:FlatVersusCurved}, Eq. \ref{eq:LDAsimplify} does not hold. For this case Snell's law can not (easily) be applied in order to transform $\theta_w$ to $\theta_a$. A prerequisite of calculating the correct velocity is therefore the knowledge of $\theta_w$ as a function of gap-position.

Most commonly the calculation of the velocity is implemented in the supplied software and implicitly assumes Eq. \ref{eq:LDAsimplify} to hold. For the case of curved interfaces this equation does not hold, and therefore the measured velocity has to be corrected by multiplying it with a correction factor $C_\theta$:
\begin{equation}
 C_\theta = \frac{u_{\phi,\text{real}}}{u_{\phi,\text{measured}}} =\frac{n_a sin(\theta_a/2)}{n_w sin(\theta_w/2)}, \label{eq:thetacorrection}
\end{equation}
where $a$ subscripts denote quantities in air, and $w$ in water, see also Fig. \ref{fig:PlanesAndAngle}.

\section{Solutions}
The problem at hand is predominantly solved by mounting prisms (see \textit{e.g.} \cite{CastelloPHD}) to the outer cylinder of the TC apparatus, or by putting the entire apparatus inside a liquid bath (see \textit{e.g.} \cite{Sun2005}) with flat windows. The latter having two purposes: the liquid bath can act as a coolant and match the refractive index of the working fluid. In this way the beams travel through the outer cylinder with less deflection; this solution is, however, not perfect because of the finite thickness of the outer cylinder. Matching the refractive indices of the working fluid, the liquid bath, and the outer cylinder does solve the issue, but becomes cumbersome for large scale devices, or impossible if the studied fluid is a gas. The use of prisms is tantamount to the use of a liquid bath, and is also unable to fully correct for the problem. Furthermore, applying prisms is technically demanding once the outside is in motion. Theoretically one can derive the trajectories of the laser light. Ref. \cite{zhang2004} derives these trajectories and even finds simplifications for the found solutions. This analysis is, however, not complete, as it only considers 1 plane of refraction; the inside of the outer cylinder. In our application our cylinder is very thick compared to our measurement range; the gap of our apparatus is \unit{8}{\centi \meter} versus a cylinder of thickness \unit{2.5}{\centi \meter}. Here we will consider both interfaces, and calculate the trajectories by using a ray-tracer, and we will show that taking into account both interfaces is \emph{crucial} for our experiments. Utilizing a ray-tracer has several advantages compared to a theoretical derivation for the present experimental setup (\emph{e.g.} Ref. \cite{zhang2004}): a theoretical derivation becomes cumbersome if one tries to find a formula after more than one refraction.  The \tttc \ system will be equipped with multiple outer cylinders to alter the gap-width, a ray-tracer is then more generalized and is able to handle multiple interfaces. The next section describes the use of a ray-tracer in order to account for the effects of the curved interfaces.

\section{Ray-tracer}
A 3D ray-tracer is built in order to calculate two parameters: the angle $\theta_w$ for the green beam pair in the horizontal plane, and the position of the crossings of the blue and green beam pairs in the vertical and horizontal plane, respectively. For the blue beam pair (in the axial-radial plane), the curvature effects does not affect the flow velocity measurements, and therefore Eq. \ref{eq:LDAsimplify} is applicable; the axial velocity does not require correction. The azimuthal velocity, however, does need correction. 

The ray-tracer is based on simple principles: starting at point $\mathbf p_i$ with direction $\mathbf k_i$, it checks which interfaces are hit for some  $t>0$ at position $\mathbf p_i + t \mathbf k_i$. The next point in the ray-trace can be defined from the interface that is hit first: $\mathbf p_{i+1} = \mathbf p_i + t_{\text{min}} \mathbf k_i$. The normal of this interface is calculated at position $\mathbf p_{i+1}$, and is denoted $\mathbf{\hat{s}}$, where the hat means the vector has unit length. For the case of reflection the new direction is given by:
\begin{equation}
 \mathbf{\hat{k}_{i+1}} = \mathbf{\hat k_i} - 2 (\mathbf{\hat{k}_i} \cdot \mathbf{\hat{s}}) \mathbf{\hat{s}}.
\end{equation}
For the case of refraction, Snell's law:
\begin{equation}
n_i (\mathbf{\hat{k}_i} \times \mathbf{\hat{s}}) = n_{i+1} (\mathbf{\hat{k}_{i+1}} \times \mathbf{\hat{s}}),
\end{equation}
is solved for $\mathbf{k_{i+1}}$ under the constraint that it has unit length and in the plane spanned by $\mathbf{\hat{s}}$ and $\mathbf{\hat{k_i}}$. This can be implemented without the use of trigonometric functions, which can be troublesome in certain fringe cases, see the appendix for more details.

Once the new position and new direction are found, the algorithm can be repeated until it exits the apparatus, or until it is absorbed by a surface.

This algorithm has been applied to the geometry of our LDA and TC setup. Our focal length $f =\unit{0.5}{\meter}$, our beam separation is \unit{76}{\milli \meter} for the green beams, and \unit{73}{\milli \meter} for the blue beams. The optical geometry of  the TC apparatus used in the present work \cite{gil10b} can be characterized by 3 radii and 3 refractive indices, see  Fig. \ref{fig:TC_Geometry} and \ref{fig:PlanesAndAngle} and Table \ref{tab:radii}.

\begin{table}[ht!]
 \begin{center}
  \begin{tabular}{| l |  c | l  l | }
  \hline
  Parameter								&		Symbol		&		Value & \\
  \hline
  Radius inner cylinder 				&	 	$r_i$ 		& 		0.20 				& \onlyunit{\meter} \\
  Inner-radius outer cylinder 	&		$r_o$ 		& 		0.28 				& \onlyunit{\meter} \\
  Outer-radius outer cylinder 	&		$r_e$		&		0.305				& \onlyunit{\meter} \\
  Refractive index PMMA			&		$n_{\text{{\tiny{PMMA}}}}$	&	1.49		& \onlyunit{\meter} \\
  Refractive index water			&		$n_{\text{{\tiny{water}}}}$	&	1.333		& \onlyunit{\meter} \\  
  \hline
  \end{tabular}
  \caption{Parameters describing the optical geometry of the presently used Taylor-Couette apparatus: the \tttc \ \cite{gil10b}.}
  \label{tab:radii}
  \end{center}
\end{table}

\subsection{Shift of focal position}
\begin{figure}[ht!]
 \begin{center}
  \includegraphics{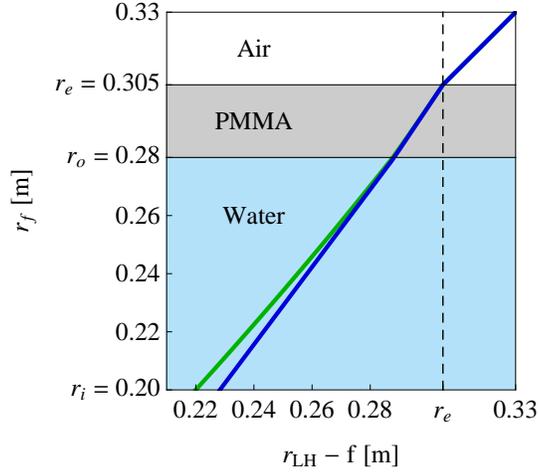}
  \caption{The position of the two foci as a function of the position of the laser-head. Colors are in accordance with Fig. \ref{fig:PlanesAndAngle}. The trajectory of the blue focus can be described by a piece-wise linear function, while for the green focus it deviates from a linear function due to the curvature of the interfaces. The foci diverge once $r_{\text{LH}}-f<r_e$.}
  \label{fig:FocalPosition}
 \end{center}
\end{figure}

In this section the location of the foci (\textit{i.e.} the measurement position) for both pairs of beams are calculated. The position of the laser-head $r_\text{LH}$ is varied, see Fig. \ref{fig:PlanesAndAngle}. The focal position of the undisturbed beams is given by $r_\text{LH} - f$. For each beam-pair the focal position $r_f$ is calculated as a function of $r_\text{LH}$, see Figs. \ref{fig:PlanesAndAngle} and \ref{fig:FocalPosition}.

\begin{figure}[ht!]
 \begin{center}
  \includegraphics{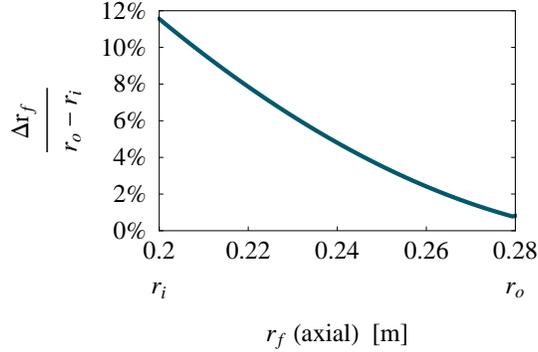}
  \caption{The separation between the measurement positions is normalized with the gap width and plotted versus the focal position of the `axial' beams. Near the inner cylinder the distance between the two foci is the highest and constitutes more than 11\% of the gap. Due to the thickness of the outer cylinder there is non-zero distance between the two foci when focused at $r=r_o$.}
  \label{fig:FocalPositionDelta}
 \end{center}
\end{figure}

If the focus is outside the apparatus, \emph{i.e.} $r_\text{LH} -f > r_e$, the focal position is given by $r_f = r_\text{LH} -f$. If the laser head is moved inward (decreasing $r_\text{LH}-f$) the beams will first hit the outer cylinder at $r_f = r_e = r_\text{LH} -f$. Moving the laser head further inward will cause the foci to lie inside the PMMA, and moving even further, inside the water. The blue beams are in the azimuthal-radial plane and refract differently from beams travelling in the axial-radial plane. The focus of the blue beams hits the inner cylinder ($r_f = r_i$) at $r_\text{LH}-f \approx \unit{0.228}{\meter}$, while the focus of the green beams hits the inner cylinder at $r_\text{LH}-f \approx \unit{0.22}{\meter}$. The distance between the two foci as a function of the radial position is depicted in Fig. \ref{fig:FocalPositionDelta}. 

A pronounced shift of the focal positions is observed, see Fig. \ref{fig:FocalPositionDelta}; the foci never coincide, and the maximum separation is 11.5\% of the gap. The effects due to curved interfaces can therefore not be neglected. Furthermore, note that the foci do not even coincide at $r=r_o$; this is due to the finite thickness of the outer cylinder, it is therefore necessary to consider both interfaces in the analysis. 

\subsection{Beam angle correction}
\begin{figure}[ht!]
 \begin{center}
  \includegraphics{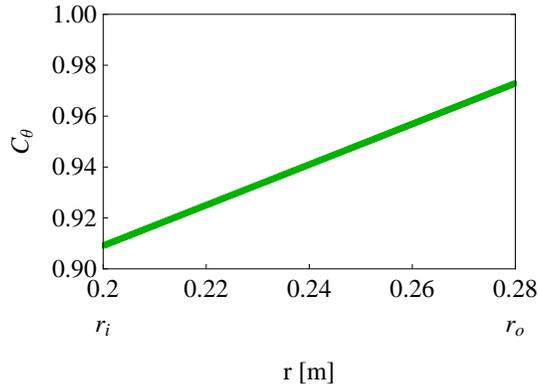}
  \caption{Correction factor $C_\theta$ (see eq. \ref{eq:thetacorrection}) as a function of the radial position $r$. The azimuthal velocity $u_\phi$ has to be corrected: $u_{\phi,\text{real}} = C_\theta (r) u_{\phi,\text{measured}}$.}
  \label{fig:ThetaCorrection}
 \end{center}
\end{figure}

For the case of flat surfaces (the axial-radial plane) the velocity calculated by the supplier's software does not have to be corrected. Equation \ref{eq:LDAsimplify}, however, does not hold in the azimuthal-radial plane due to the curved surfaces, and the velocity has to be corrected by multiplication with $C_\theta$, see Eq. \ref{eq:thetacorrection}. The refractive indices are known, and $\theta_a$ can be found from the focal length and the beam separation. $\theta_w$ can be found by calculating the angle between the focussing rays, see Figs. \ref{fig:PlanesAndAngle} and \ref{fig:FlatVersusCurved}. The correction factor can then be calculated, see Fig. \ref{fig:ThetaCorrection}.

The velocity has to be corrected by 3\% to 9\%, depending on the radial position. Note that there is a finite correction (about 3\% for the optical geometry of the current apparatus) at the water-PMMA interface at $r=r_o$, this is due to the non-zero thickness of the outer cylinder and taking into account both interfaces of the outer cylinder in the analysis. 

\section{Experimental verification}
The results obtained from the ray-tracer can be verified by measuring a known flow state. The temperature of the TC system is kept at $20^\circ$C with water as the working fluid. Dantec polyamid seed particles ($r_{\text{seed}}=\unit{2.5}{\micro \meter}$) with a density of $\unit{1.03\cdot 10^3}{\kilo \gram \per \meter^3}$ are used. One can estimate the minimum velocity difference $\Delta v = |v_{seed} - v_{fluid}|$ between a particle $v_{seed}$ and its surrounding fluid $v_{fluid}$ needed for the drag force $F_{drag} = 6 \pi \mu r_{seed} \Delta v$ to outweigh the centrifugal force $F_{cent}(r) = \frac 43 \pi {r_{seed}}^3 \left( \rho_{seed} - \rho_{fluid}\right) \frac{v^2}{r}$. A typical velocity in the middle of the gap ($r = 0.24$ m) is $v = \unit{5}{\meter \per \second}$, combining with the density and viscosity of water around $20^\circ$C, resulting in $\Delta v \approx \unit{4\times 10^{-6}}{\meter \per \second}$. This is several orders of magnitude smaller than the typical velocity fluctuation inside the TC-gap of order $10^{-1}$ m/s and hence centrifugal forces on the seeding particles are negligible. 

\begin{figure}[ht!]
 \begin{center}
  \includegraphics{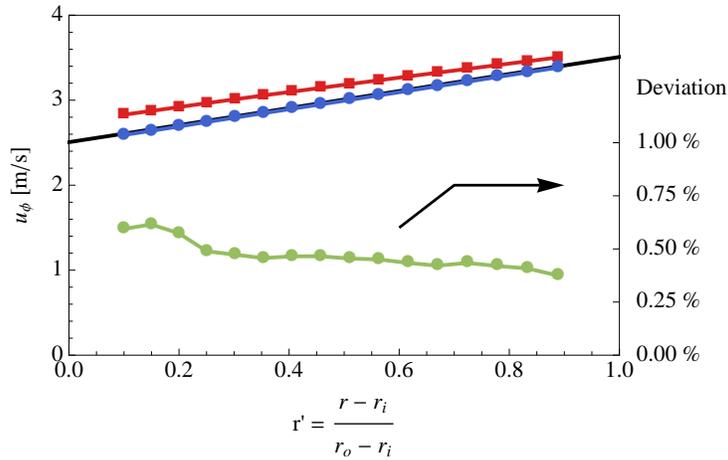}
  \caption{In red squares the \emph{uncorrected} azimuthal velocity as a function of radial position is shown, and in blue dots the \emph{corrected} azimuthal velocity. The black solid line is the theoretical flow profile $u_\phi = \omega r$, the deviation from this profile is plotted with green dots, the corresponding scale is on the right. The theoretical and corrected measured profiles are found to be in good accordance.}
  \label{fig:verification}
 \end{center}
\end{figure}

For TC flow a stable and well-known flow state is solid body rotation; the inner and outer cylinder are both rotated at a fixed speed $\omega$. After sufficient waiting the fluid will have a velocity $u_\phi = \omega r$, and $u_z = u_r = 0$. The experiment has been performed for three rotation rates $\omega$ ($\omega/2\pi = \unit{1}{\hertz}$, $\unit{2}{\hertz}$, and $\unit{4}{\hertz}$), where the azimuthal velocity has been measured at several radial positions and at mid height. Similar results were found for all three cases, Fig. \ref{fig:verification} shows the results for the case of $\omega/(2 \pi) = \unit{2}{\hertz}$.

The measured velocities are shown in red squares, after applying the beam angle correction the data points (blue circles) are found to agree with the theoretical flow profile within 0.75\%. Any remaining deviation can be due to \emph{e.g.} optical misalignment or imperfection, spatially inhomogeneous refractive index in the working fluid, or noise created by the amplification and digitalisation of the optical signal.

\section{Application}
\begin{figure}[ht!]
 \begin{center}
  \includegraphics{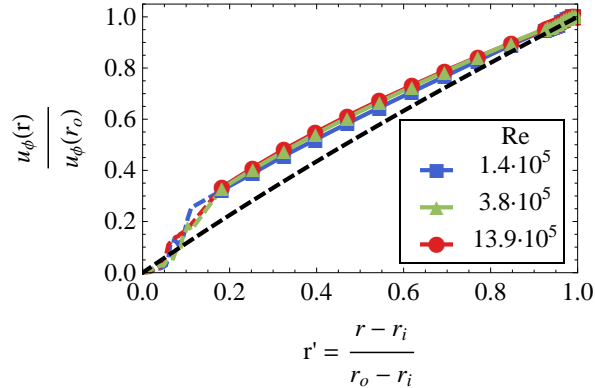}
  \caption{For three rotation rates the azimuthal velocity is measured along the gap. This velocity is normalized by the driving velocity in order to collapse the data, and compared to the laminar velocity profile for infinite aspect ratio (dashed line). The profile is found to be nearly independent of the Reynolds number.}
  \label{fig:PureOuterRotation}
 \end{center}
\end{figure}

Here the results are shown for the measurement of the azimuthal velocity profile for the case that the inner cylinder is stationary and the outer cylinder is rotating. This case has been studied before \cite{wen33,col65a}, and because the flow is laminar, even for high Reynolds number, any perturbation due to a measuring probe is likely to survive a full revolution. To accurately obtain the speed of the (undisturbed) flow, it has to be measured non-intrusively. Fig. \ref{fig:PureOuterRotation} shows the results of three experiments having varying Reynolds number $\left( Re= \omega_o r_o (r_o-r_i) / \nu \right)$.

For three rotation rates the obtained profiles are found to collapse over a decade of Reynolds number, but deviate from the laminar velocity profile for infinite aspect-ratio. Due to a finite aspect ratio of the setup the presence of the end plates will create a secondary flow, modifying the azimuthal velocity profile. The velocity close to the inner cylinder is currently unattainable by LDA due to reflections coming of the inner cylinder surface. These reflections create spurious, unreliable data and therefore the profiles are shown dashed in that region. The present system will, in the future, be equipped with a transparent inner cylinder, which will significantly reduce the reflections. 

\section{Conclusion}
In order to measure correct velocities inside a TC apparatus using LDA, one has to correct for the effects of the curved interfaces. Not only do the positions of the measurement volumes depend non-trivially on the position of the laser head, but also the measured velocity has to be corrected. A ray tracer has been used in order to calculate the position of the foci, but also to calculate the correction factor $C_\theta$ as a function of radial position. We showed that for our application it is crucial to take into account both interfaces of the outer cylinder. The measurement positions do not coincide and the velocity has to be corrected even at $r=r_o$. Our ray-tracer is verified by measuring the velocity for the case of solid body rotation; good agreement with the theoretical prediction has been found. For pure outer cylinder rotation it is found that the velocity deviates from the laminar velocity profile for infinite aspect ratio due to the presence of the end plates.

\section*{Acknowledgements}
This work was financially supported by technology foundation STW in The Netherlands. Stimulating discussions with Jan Abshagen, Ren\'e Delfos, Detlef Lohse, Gerd Pfister, Florent Ravelet, and Jerry Westerweel are acknowledged. We also acknowledge the help of G.W. Bruggert, M. Bos and B. Benschop for their technical support. 

\appendix
\section*{Appendix}

\begin{figure}[ht!]
 \begin{center}
  \includegraphics{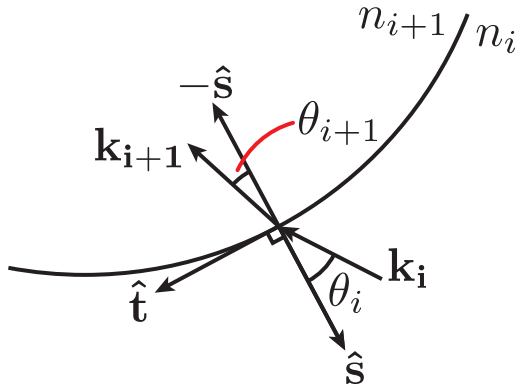}
  \caption{Sketch of the rays in the plane of refraction.}
  \label{fig:appendix}
 \end{center}
\end{figure}

For each interface one has to solve Snell's law $n_i (\mathbf{\hat k_i} \times \mathbf{\hat s})=n_{i+1} (\mathbf{\hat k_{i+1}} \times \mathbf{\hat s})$ for $\mathbf{k_{i+1}}$. From the vector equation one can see that the solution is not unique; but the solution is unique if we require the solution to be in the plane spanned by $\mathbf{\hat{s}}$ and $\mathbf{\hat{k_i}}$ (called the plane of refraction). In this plane the solution can be rewritten as: $n_i \sin(\theta_i) = n_{i+1} \sin(\theta_{i+1})$. $\mathbf{k_{i+1}}$ can be decomposed in to two parts; one in the direction of $\mathbf{\hat{s}}$ and one that is tangential to the interface and in the plane of refraction, denoted $\mathbf{\hat t}$, see figure \ref{fig:appendix}. By simple geometry one can find that $\mathbf{k_{i+1}} = \sin(\theta_{i+1}) \mathbf{\hat t} + \cos(\theta_{i+1}) (-\mathbf{\hat{s}})$. The direction of $\mathbf t$ can be found by subtracting the normal component of the incident ray: $\mathbf t = \mathbf{\hat{k_i}}-(\mathbf{\hat{k_i}}\cdot\mathbf{\hat{s}})\mathbf{\hat{s}}$. From the definition of the cross product ($\left| \mathbf a \times \mathbf b \right| = |\mathbf a||\mathbf b|\sin(\alpha)$, where $\alpha$ is the angle between vectors $\mathbf a$ and $\mathbf b$) and Snell's law, one can derive that $\sin(\theta_{i+1}) = \frac{n_1}{n_2} \left| \mathbf{\hat k_i} \times \mathbf{\hat s} \right|$, and using the Pythagorean identity one writes $\cos(\theta_{i+1})=\sqrt{1-\sin(\theta_{i+1})^2}$. Note that one does not need to calculate the angles $\theta_i$ or $\theta_{i+1}$, it is only necessary to know the sine and or cosine of the angles. The direction for the outgoing ray can now be found by substituting all the values: $\mathbf{k_{i+1}} = \sin(\theta_{i+1}) \mathbf{\hat t} + \cos(\theta_{i+1}) (-\mathbf{\hat{s}})$.

\end{document}